# A tunable and versatile 28nm FD-SOI crossbar output circuit for low power analog SNN inference with eNVM synapses

Joao Henrique Quintino Palhares[1,2,3,4], Yann Beilliard[3,4], Jury Sandrini[1], Franck Arnaud[1], Kevin Garello[2], Guillaume Prenat[2], Lorena Anghel[2], Fabien Alibart[3,4,5], Dominique Drouin[3,4], Philippe Galy[1]

[1]STMicroelectronics, 850 rue Jean Monnet, 38920 Crolles, France.
[2]Univ. Grenoble Alpes, CEA, CNRS, Grenoble INP, SPINTEC, 38000 Grenoble, France.
[3]Institut Interdisciplinaire d'Innovations Technologiques (3IT); Université de Sherbrooke; 3000 Bd de l'Université, Sherbrooke, Québec, Canada
[4]Laboratoire Nanotechnologies Nanosystèmes (LN2); CNRS IRL 3463, Université de Sherbrooke; 3000 Bd de l'Université Université, Sherbrooke, Québec, Canada
[5]Institute of Electronics, Microelectronics and Nanotechnology (IEMN), Université de Lille, Villeneuve d'Ascq 59650, France

Email: joaohenrique.quintinopalhares@st.com
Email: philippe.galy@st.com

## 1. Abstract

In this work we report a study and a co-design methodology of an analog SNN crossbar output circuit designed in a 28nm FD-SOI technology node that comprises a tunable current attenuator and a leak-integrate and fire neurons that would enable the integration of emerging non-volatile memories (eNVMs) for synaptic arrays based on various technologies including phase change (PCRAM), oxide-based (OxRAM), spin transfer and spin orbit torque magnetic memories (STT, SOT-MRAM). Circuit SPICE simulation results and eNVM experimental data are used to showcase and estimate the neurons fan-in for each type of eNVM considering the technology constraints and design trade-offs that set its limits such as membrane capacitance and supply voltage, etc.

Keywords; UTBB 28nm FD-SOI, Analog SNN, Analog eNVM, eNVM integration.

## 2. Introduction

Spiking neural networks (SNN) based on emerging non-volatile memory (eNVM) crossbars are promising in-memory computing components that exhibit outstanding capabilities for low power artificial intelligence at the edge. However, the co-integration of eNVMs synaptic arrays with 28nm ultra-thin body and buried oxide fully-depleted silicon-on-insulator (UTBB-FDSOI) technology node remains a challenge. In analog spiking neural network (SNN), input neurons are interconnected with output neurons through one-resistor-one-transistor (1T1R) synapses and the computation is done through voltage spikes converted into current by synaptic weights [1]. The neurons accumulate the spikes up to a predefined threshold then an output spike is generated. The neuron capability to distinguish and to accommodate a massive number of synapses and input spikes is directly related to the voltage swing up to the firing threshold of the neuron. This is mostly determined by the membrane capacitance, the net number of synaptic charges and the threshold of the low power neuron [2]. The number of synapses the neuron can accommodate is called the crossbar fan-in of a SNN implemented in hardware using eNVMs synapses. Therefore, to increase fan-in, it is mandatory either to increase the membrane capacitance or to reduce the synaptic excitation current. Another alternative is to increase the firing threshold of the neuron, which is normally limited by the circuit power supply voltage and CMOS technology node. In addition, it should be noted that the size of the neuron is mostly determined by the membrane capacitor ($C_{mem}$), and so the capacitance choice is restricted to the silicon area available. In turn, a promising approach to solve these issues hindering the scale up of SNN hardwares consists in using current attenuators to reduce the synaptic current [2,3]. In this work we present a comparison analysis of an eNVM-synapse oriented crossbar output circuit designed on 28nm FD-SOI technology, and a co-design methodology to keep track of the fan-in of the analog neuron for different eNVM technologies and circuit design constraints. The circuit comprises current attenuators and analog spiking neurons. The eNVMs technologies considered in the analyses are PCRAM, OxRAM and STT, SOT-MRAM).

## 3. Hardware Design

The 28nm FD-SOI low power analog Leaky integrate-and-fire (LIF) neurons were designed based on previous team research [4] and simulated using thick oxide high k metal gate, regular $V_t$ grade transistors and four different poly n-well membrane capacitances: 864.5fF, 86.4fF, 8.6fF and 1.4fF, in which the latter is rather the native lumped parasitic capacitance at the neuron's input node ($P_2+N_2$), see Fig. 1a. The neuron area overhead is shown in Fig.1b-c.

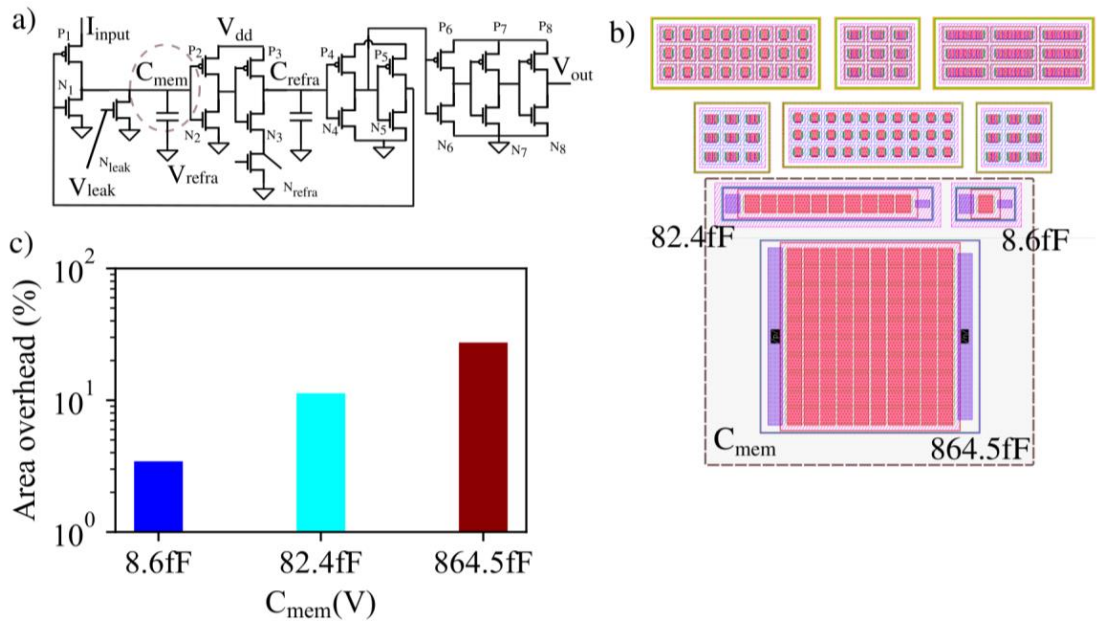

**Figure 1:** a) 28nm FD-SOI post-synaptic analog neuron circuit b) layout and c) membrane capacitor area overhead.

The transfer characteristics and the neuron time constant are presented in Fig.2a-b. The neuron firing frequency corresponds to constant input synaptic current excitations ($I_{input}$) and the time constant was acquired for different $V_{leak}$ from 0.1 to 0.4V.

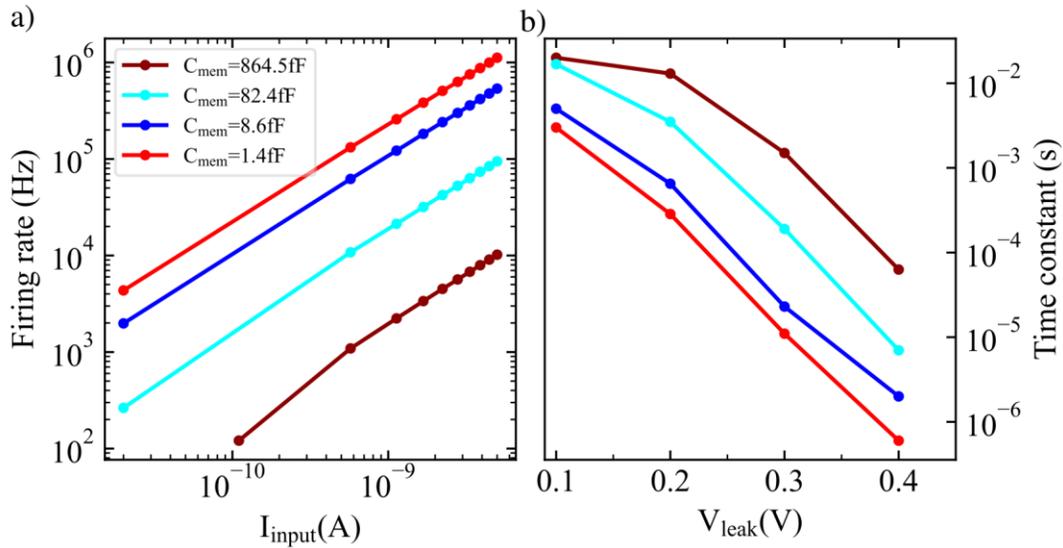

**Figure 2:** (a) 28 nm FD-SOI analog neuron circuit SPICE simulation results at room temperature with typical process corner for different $C_{mem}$: Firing frequency vs. synaptic excitation current and b) time constant vs voltage-controlled leakage current at room temperature.

To operate the neurons within this input current range starting from 20pA a 28nm FD-SOI current attenuator was designed using the same transistor grade (see Fig.3). The current attenuator topology is based on reference [3] and adapted to 28nm FD-SOI technology. The current attenuator reduces the input current of the circuit by transferring the voltage drop in the transistor $N_9$ in series with the eNVM synapses, to the input of a differential transconductance amplifier.

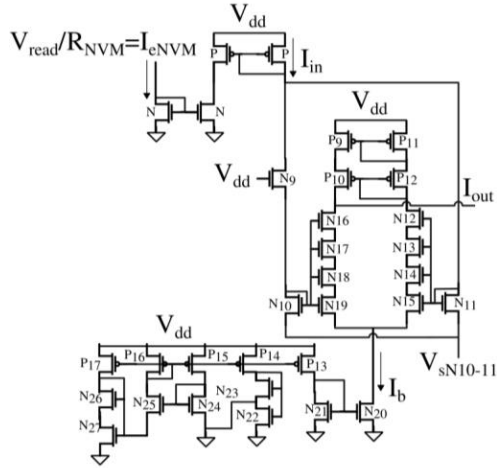

**Figure 3:** 28nm FD-SOI current attenuator circuit.

The output current $I_{out}$ is limited by a very small biasing current $I_b$ and is proportional to the voltage drop of transistor $N_9$ and then to $I_{in}$ and the eNVM synaptic current ($I_{eNVM}$). Thus, the reduced version of the synaptic current is [5]:

$$I_{out}=I_{dsN16}-I_{dsN12}=I_b\text{Tanh}\left(\frac{K_n}{2U_t}\left(R_{N9}\cdot\frac{\text{Iin}}{2}\right)\right) \quad (1)$$

$K_n$ is the subthreshold slope and $U_t=KT/q$ the thermal voltage and $R_{N9}$ is the resistance of transistor $N_9$. The resulting $I_{in}/I_{out}$ ratio is the current scaling down factor (SDF). An important aspect is that the SDF should be continuous in the crossbar read out range so as not to distort the network response. The neurons receive input spikes from several pre-synaptic neurons. Note that a SNNs are accessed in a very sparse way, therefore the row activation frequency is very small [1]. Thus, one can consider this readout range of one average eNVM, disregarding variations far beyond the edge of the reading margin of the eNVMs, as shown in Fig.4.

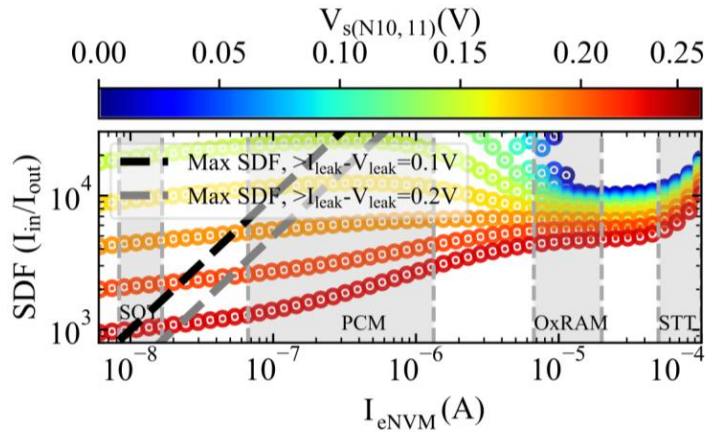

**Figure 4:** 28nm FD-SOI current attenuator circuit spice simulation results of SDF for different $V_{sN10,11}$ bias at room temperature with typical corner. The regions of readings from each eNVM marked in the shaded portions, together with the line marking the maximum SDF due to the leakage current of the neuron, delimit the appropriate continuous attenuation bands.

The resistance ranges of each eNVM considered in this analysis are based on experiments at room temperature with phase change memories (PCM) provided by STMicroelectronics, and reference values for oxide-based memories (OxRAM) provided by 3IT[6], and spin transfer and spin orbit torque magnetic memories (STT, SOT) from reference[7]. Another setting condition is that the resulting current after attenuation must be higher than the leakage of the neuron, so there is a maximum SDF for each eNVM represented by the dashed line in Fig.4. Another setup constraint that must be obeyed for equation (1) to be valid and to obtain an approximately linear current conversion with respect to $I_{eNVM}$ is the saturation of $N_{20}$. This is satisfied for $V_{dN20}>4U_t$ or $4U_t<(K_n(V_{dN10}+V_{dN11})-K_nN_{20}V_{gN20})/2$. To reach this condition $V_{sN10,11}$ must be adjusted and so $V_{dN10,11}$ of diode connected $N_{10,11}$ at the cost of decreasing SDF due to the reduction in $I_{in}$ for the same $I_{eNVM}$. Thus, for each eNVM the current attenuator must be set to a different SDF. Based on this framework, a silicon test chip (see Fig.5) with the crossbar output circuit that comprises the current attenuator and the analog neuron is under development.

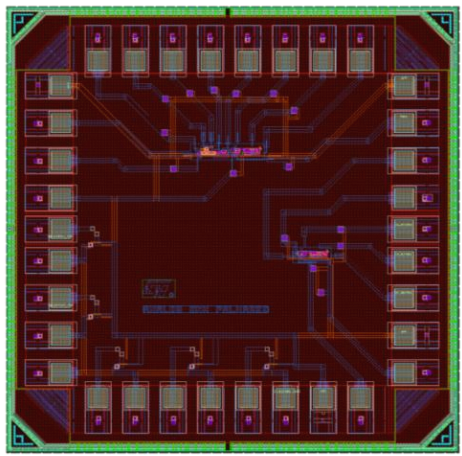

**Figure 5:** (a) 28 nm FD-SOI crossbar output circuit, 1mm$^2$ test chip.

## 4. Scalability Abacus

According to reference [2], a large-scale network would accommodate between $10^2$ and $10^3$ spikes while a small-scale network between 10 and $10^2$ input spikes. Thus, using the experimental data from the eNVMs and simulation results from the neuron and current attenuator, the approach used to determine the minimum fan-in was (i) to estimate the increase in the membrane potential $\Delta V_{mem}$ after receiving a spike from a synapse in low resistance state (LRS) - the most demanding scenario-, (ii) to divide the threshold voltage of the neuron by the average increase in membrane potential per synapse, and (iii) to estimate the amount of average synapses that the neuron is able to accommodate before firing, which corresponds to the minimum fan-in. The voltage increment in the membrane capacitor per average synaptic stimulus is:

$$dv = \frac{(\frac{V_{read}}{R_{LRS}} \cdot \frac{1}{SDF} - I_{leak}) \cdot t_{pulse\_width}}{C_{mem}}$$

$$dv = \frac{(\frac{V_{read}}{R_{LRS}} - I_{leak}) \cdot t_{pulse\_width}}{C_{mem}}$$

$$dv = \frac{(I_{input} - I_{leak}) \cdot t_{pulse\_width}}{C_{mem}}$$

$$\Delta V_{mem} = \frac{(\frac{V_{read}}{R_{LRS}} \cdot \frac{1}{SDF} - I_{leak}) \cdot t_{pulse\_width}}{C_{mem}} \quad (2)$$

$$\text{Fan-in} = \frac{V_{th}}{dv} = \frac{\frac{V_{dd}}{2}}{dv} \quad (3)$$

Therefore, according to our framework, to increase the fan-in of the post-synaptic neuron, one needs to increase the SDF, which has a limit based on each eNVM and $V_{leak}$. It is also necessary to increase Vth, that depends on voltage supply, or decrease the pulse width and amplitude. The fan-in for each eNVM and neuron capacitance are shown in Fig. 6a,b.

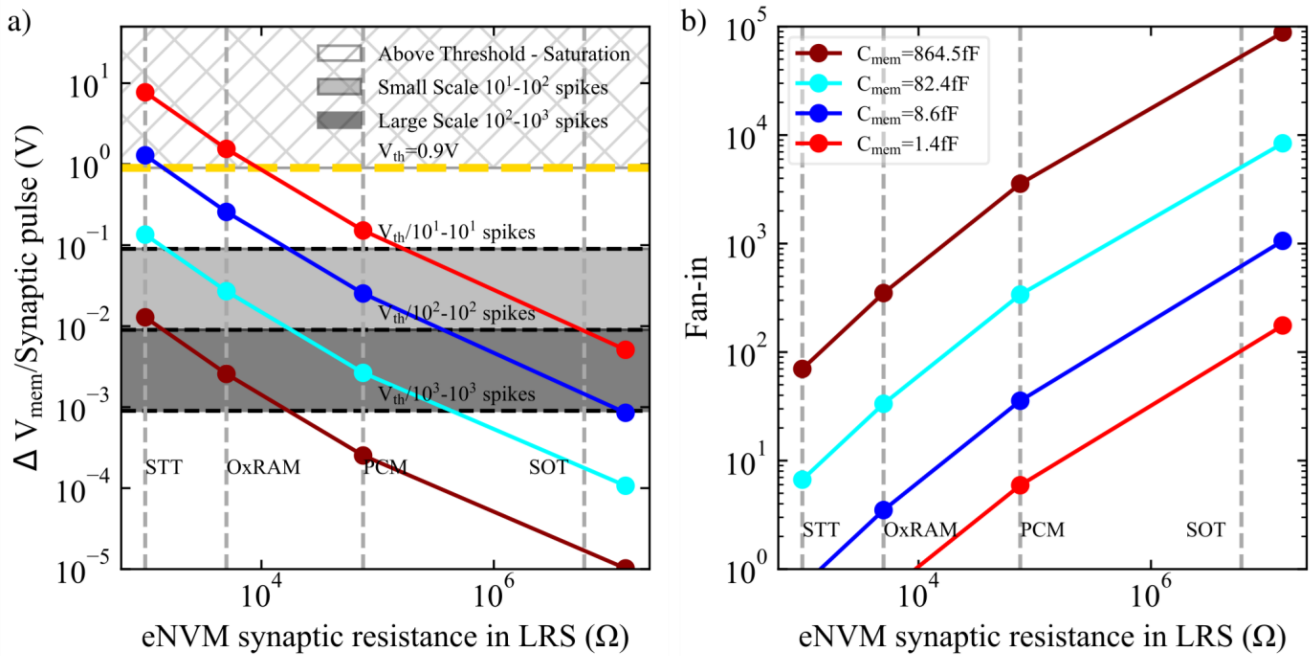

**figure 6:** a) Output neuron ΔVmem and b) fan-in for each eNVM synapses and Cmem assuming pulse width of 1μs, 0.1V, Vleak of 0.1V and SDFs of 9000, 9000, 6000 and 1000 for STT, OxRAM, PCM and SOT, respectively. The resistance range corresponding to each eNVM is indicated by the dashed line.

The crossbar column could accommodate at least approximately 3560 PCMs, 88200 SOTs and 350 OxRAM synapses in LRS for a membrane capacitance of 864.5fF, pulse width of 1μs, $V_{leak}$ of 0.1V, Vth of 0.9V and SDF of $6.10^3$, $10^3$ and $9.10^3$, respectively. STT fits at least a small-scale demonstrator using regular Vt thick oxide transistors, capacitances up to 864.5fF and current reduction of $9.10^3$. One can conclude through this careful analysis that high resistance eNVMs are most likely to be implemented in large-scale crossbars. In addition, besides the fan-in, multilevel programming is paramount in analog SNNs, and this is another challenge aspect that impacts cell size, the crossbar programming, and the implementation strategy. Multilevel programming, memory non-idealities, drift and dispersion will be addressed in a future work.

## 5. Conclusion

We demonstrated a co-design methodology to estimate the minimum fan-in of a 28nm FD-SOI analog SNN crossbar output circuit for different eNVM synapses (PCM, OxRAM, STT and SOT MRAM). It was shown that many factors may influence the fan-in of the network such as the power supply voltage, the resistance range of the eNVMs, the reading pulse, the leakage and membrane capacitance of the neuron, and the operating conditions of the current attenuator. Despite all these factors, the most important determinants of scalability are the resistance range of the eNVM, the membrane capacitance and the current scaling down factor of the current attenuator. To our knowledge there is no fair comparison analysis of different eNVM technologies that jointly address the limits of application of analog SNNs with FD-SOI 28nm output circuits. Therefore, our results and agnostic co-design methodology pave the way for a deeper analysis of integration of eNVMs into analog SNN on FD-SOI 28nm technology and more advanced nodes.

**Acknowledgements**

This work was financially supported by STMicroelectronics SA and the French CIFRE Program (Conventions Industrielles pour la Formation par la Recherche). We also acknowledge the SPINTEC IC design team and the Laboratoire Nanotechnologies et Nanosystèmes (LN2)/International Research Laboratory.